\documentclass[twocolumn]{aastex631}

\usepackage{amsmath}
\usepackage{float}
\usepackage{scalerel}
\usepackage{lipsum}
\usepackage {hyperref}
\usepackage{rotating, graphicx}

\shorttitle{A Statistical Analysis of Galactic SNRs}
\shortauthors{Ranasinghe \& Leahy.}
\graphicspath{{./}{figures/}}

\begin{document}

\title{A Statistical Analysis of  Galactic Radio Supernova Remnants}

\author[0000-0002-9559-3827]{S. Ranasinghe}
\affiliation{Department of Physics and Astronomy, University of Calgary, 2500 University Dr NW, Calgary, AB, T2N 1N4, Canada}

\author[0000-0002-4814-958X]{D. Leahy}
\affiliation{Department of Physics and Astronomy, University of Calgary, 2500 University Dr NW, Calgary, AB, T2N 1N4, Canada}


\begin{abstract}
 We present an revised table of 390 Galactic radio  supernova remnants (SNRs) and their basic parameters. Statistical analyses are performed on SNR diameters, ages, spectral indices, Galactic heights and spherical symmetries. Furthermore, the accuracy of distances estimated using the $\Sigma$-D relation is examined. The arithmetic mean of the Galactic SNR diameters is $30.5$ pc with standard error $1.7$ pc and standard deviation $25.4$ pc. The geometric mean and geometric standard deviation factor of Galactic SNR diameters is $21.9$ pc and  $2.4$, respectively. We estimate ages of 97 SNRs and find that the supernova (SN) birth rate to be lower than, but within $2\sigma$ of currently accepted values for SN birth rate. The mean spectral index of shell-type SNRs is $-0.51 \pm 0.01$ and no correlations are found between spectral indices and the SNR parameters of molecular cloud (MC) association, SN type, diameter, Galactic height and surface brightness. The Galactic height distribution of SNRs is best described by an exponential distribution with a scale height of $48 \pm 4$ pc. The spherical symmetry measured by the ovality of radio SNRs is not correlated to any other SNR parameters considered here or to explosion type.
\end{abstract}



\section{Introduction} \label{sec:intro}

For their role in Galactic and interstellar medium (ISM) evolution, the study of supernovae (SNe) and their remnants (SNRs) is vital. In order to better understand SNRs, their basic parameters such as distance, age, spectral index etc., as well as  an accurate count are necessary. \\
\indent  \cite{2019GreenCat} gives the current total of Galactic SNRs as 294. While the predicted total number of Galactic SNRs is a few thousand (\cite{2018Supan}, \cite{2022RanasingheLeahy}), observations and identification of SNRs are greatly affected by selection effects. \\
\indent  Several hurdles lead to a smaller sample of Galactic SNRs ($ < 294$) with which to perform statistical studies. Out of the observed SNRs, many do not have a distance or an age because of the difficulty in estimating them. The distances to SNRs are estimated using HI absorption spectra and maser and molecular cloud (MC) associations.  When distances using these methods are unavailable, distances are often estimated using the $\Sigma$-D relation (surface brightness ($\Sigma $)- physical diameter (D) relation). The main drawback of the $\Sigma$-D relation is the large scatter in the $\Sigma$-D plane leading to distances that vary from the true distance by about an order of magnitude \citep{1991Green, 2005Greenmem}. \cite{2005Xu} presented  five methods to estimate the ages of SNRs. Only a subset of SNRs have parameters needed to estimate ages leading to even smaller samples. \\
\indent While the Galactic SNR sample is far from complete, many authors have performed a statistical study on them (e.g., \cite{1991Green, 2005Xu}) where the the sample sizes range from 174 to 234. The sample we utilize for this work is an up-to-date list of Galactic SNRs which includes uncertain or newly identified SNRs. The $390$ sources we use in this work is at least a $\sim 60\%$ increase from previous studies. \\
\indent In Section \ref{SNRlist} we present the updated catalogue of 390 SNRs SNR with updated distances and other parameters. In Section \ref{ResandDis}, the analysis and discussion on the radio size distribution of Galactic SNRs, SNR distances and $\Sigma$-D relation, SNR ages, SNR spectral index distribution, galactic height distribution of SNRs and SNR spherical symmetry is given. Finally, the conclusions are given in Section \ref{conclusions}. 

\section{Supernova Remnant Source List and Their Parameters}  \label{SNRlist}

\indent For the source list, we use the catalogues presented by \cite{2019GreenCat}\footnote{\url{http://www.mrao.cam.ac.uk/surveys/snrs/}} and \cite{2012Ferrand}\footnote{\url{http://www.physics.umanitoba.ca/snr/SNRcat}}. Additionally,  an independent literature search was performed to search for newly identified SNRs. The catalogue presented by \cite{2019GreenCat} has 294 SNRs \footnote{Some of the SNRs in the catalogue have been removed (e.g. G$16.8-1.1$) as more information became available.}. The catalogue presented by \cite{2012Ferrand} has 383 SNRs which includes the 294 SNRs from the \cite{2019GreenCat} catalogue. The SNR list presented by \cite{2012Ferrand} includes new and uncertain SNRs. The independent literature search produced 7 newly identified SNRs bringing the total to 390 sources. However, for this study we do not include the SNR candidates (e.g. presented by \cite{2017Anderson}) as a caution  since it may add more contamination than information about real SNRs. The Galactic SNR list and their parameters are given in Table \ref{tab:1} (the table in its entirety is in the online version of the manuscript). The references for the spectral indices, MC interactions, ages and SN types are in the online version of Table \ref{tab:1}. The source numbers (Table \ref{tab:1}) first follow the \cite{2019GreenCat} catalogue ($1-294$) followed by the remaining \cite{2012Ferrand} catalogue ($295-383$) SNRs and finally the newly identified SNRs ($384-390$).\\  
\indent The angular sizes, the SNR types and the 1-GHz flux densities are from the catalogues. The radio spectral indices were found in the catalogues as well, however, performing a literature review we verified the spectral indices. For some cases where the 1-GHz flux densities were not given, we infer them from the spectral indices and  flux densities at other frequencies.\\
\indent  Of the 390 SNRs, 270 SNRs have a radio spectral index while 110 do not. Here the radio spectral index ($\alpha$) is defined as $ S_{\nu} \propto \nu^{\alpha}$, where $S_{\nu}$ is the flux density at frequency $\nu $. However, the radio spectrum  for some SNRs is not well described by a simple power-law, either because the radio spectrum is described by multiple power-laws or because of the variations of the spectral index across the face of the SNR  \citep{2019GreenCat}. There are 7 such SNRs in the sample where we denote the varying spectral index as ``varies''. Furthermore, there are 3 SNRs with spectral indices but the values are uncertain  (denoted with a `?').  The 1-GHz flux densities are given for 183 SNRs and uncertain flux densities (denoted with a `?') in Table \ref{tab:1}  are given for 108 SNRs (none for 99 SNRs), bringing the total number of SNRs with flux densities to 291. \\
\indent The distances for this study are literature estimations  (see \cite{2022RanasingheLeahy}; the methods of the distance estimations and references therein). We have added 6 more SNRs (uncertain) to the list with distances bringing the total number of sources with distances to 221. \cite{2019GreenCat} removed 5 SNRs from the catalogue because \cite{2017Anderson} presented evidence that they may have been confused with HII regions. We have included these uncertain SNRs, namely G$20.4+0.1$, G$21.5-0.1$, G$23.6+0.3$, G$59.8+1.2$ and G$65.8-0.5$ in our source list. It should be noted that four of these SNRs have distances. The other 3 sources with distances that have been included in the source list are G$190.2+1.1$, G$192.8-1.1$ and G$359.9+0.0$. For the SNR (uncertain) G$359.9+0.0$, its association with Sgr A* places the SNR at the Galactic centre (GC) \citep{2006WangLu}. \\
\indent The ages of SNRs presented in Table \ref{tab:1} are literarature values except for the ones that are followed by a `*'. The ages of SNRs with a `*' were estimated using the software presented by  \cite{2017LeahyWilliams} which uses X-ray emission to obtain more accurate age estimates. There are 115 SNRs with well constrained ages and 6 SNR where the ages are given as a lower or upper limit. For this work we estimated ages for 95 SNRs. A description of the method and parameters used to estimate the ages are given in Section \ref{Ages}. \\

\section{Analysis and Discussion} \label{ResandDis}

\indent The statistical analysis and discussion of the results for the 390 SNRs listed in Table \ref{tab:1} are presented here. First, we explore the size distribution of the Galactic SNRs and the accuracy of the $\Sigma$-D Relation for the Galaxy. Next the age and spectral index distributions are investigated. Finally, the Galactic height distribution and spherical symmetry of SNRs are investigated to ascertain whether trends exist for a particular class of SNRs.  

\subsection{The Radio Size Distribution of Galactic SNRs} \label{sizedist}

\begin{figure*}
\centering
\includegraphics[width=\textwidth]{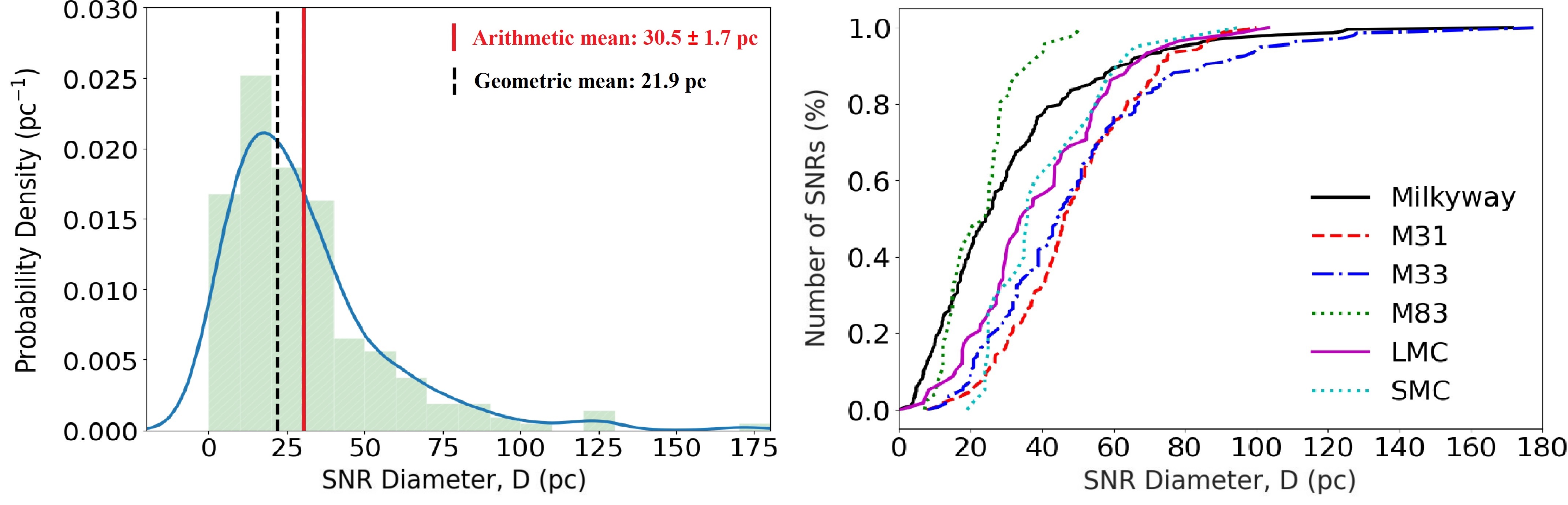}
\caption {Left: The smoothed probability density function of the SNR diameters. The red solid line is arithmetic mean ($30.5 \pm 1.7$ pc) and the vertical black dashed line is the geomatric mean of the diameter ($21.9$ pc) of the Galactic SNRs. The histogram is in the background (in green). Right: The cumulative distribution function of SNR diameters of galaxies. The black solid line- SNR diameters from this work (sample size- 214), The red dashed line- M31 from \cite{2014LeeLee} (sample size- 156), The blue dash-dot line- M33 from \cite{2010Long} (sample size- 137), green dotted line- M83 from \citep{2010Dopita} (sample size- 47), magenta solid line- LMC from \cite{2017Bozzetto} (sample size- 59) and cyan dotted line- SMC from \cite{2005FilipovicPayne} (sample size- 21).    } 
\label{fig:3}
\end{figure*}

\indent There are 221 Galactic SNRs (including uncertain SNRs) with distances. Out of the 221, 6 SNRs gives the distances as a lower or an upper limit. We exclude  G$1.4-0.1$ since its location near the GC lacks strong evidence. The radii of the remaining 214 SNRs are given in Table \ref{tab:1}. In case where the angular size of a SNR is given as the major and minor axes, the average radius (r$_{Avg}$) was estimated using r$_{Avg}$ = $ \sqrt{r_{maj}r_{min}}$, where r$_{maj}$ and r$_{min}$ are the semi-major and semi-minor axes measured in radio, respectively. For the error of the mean radius,  we adopt the error estimated using semi major axis. The probability density function (PDF) for the radio diameters is shown in Figure \ref{fig:3} (left panel).\\
\indent The statistics on the Galactic SNR diameters are given in table \ref{tab:2} and as a comparison, statistics on SNR diameters from other galaxies are included as well. The arithmetic mean of the Galactic SNR radio diameters is $30.7 \pm 1.7 $ pc. The estimated error here is the standard error, given by $\sigma_{s}/\sqrt{n}$, where $\sigma_{s}$ is the sample standard deviation and $n$ is the number of elements in the sample. The median of the sample is $24.3$ pc. The geometric mean of Galactic SNR radio diameters is $21.9$ pc. The geometric standard deviation factor ($\sigma_g$  = exp$^{\sigma(ln(D)}$) of the SNR diameter sample is 2.4. \\
\indent For M33, the mean and median SNR diameter are $\sim50$ and $44$ pc, respectively \citep{2010Long} (see Table \ref{tab:2}). \cite{2014LeeLee} gives the mean diameter of SNRs/candidates as $\sim48$ pc found in M31. The mean diameter of confirmed SNRs in the Large Magellanic Cloud (LMC)  is $39\pm4$ pc \citep{2017Bozzetto} and the Small Magellanic Cloud (SMC) SNR/candidate mean diameter is $40.8$ pc \citep{2005FilipovicPayne}. While the mean SNR diameters from M31 and M33 are consistent with each other, they are larger than that of the Milkyway SNR mean diameter. Similarly, the mean diameters of LMC and SMS are consistent with each other and larger than the Galactic mean SNR diameters. The mean diameter of SNRs in M83 is $22.7$ pc \citep{2010Dopita} and is smaller than the Milkyway mean SNR diameter. \\
\indent The geometric mean of the Galactic SNR diameters (21.9 pc) is smaller than that of the geometric mean of SNR diameters from galaxies M31, M33, LMC and SMC. As seen in Figure \ref{fig:3} left panel, the Galactic SNR diameters are skewed with a bias towards SNRs with smaller physical diameters. The geometric mean of Galactic SNR diameters is comparable to the M83 SNR diameter means. 

\begin{deluxetable*}{lrrcccccc}
\tablenum{2}
\label{tab:2}
\tablecaption{Statistics on SNR diameters of galaxies}
\tablewidth{700pt}
\tabletypesize{\small}
\tablehead{
\colhead{Galaxy} & \colhead{Sample} & \colhead{References} &\colhead{Arithmatic}  & \colhead{Standard} &  \colhead{Standard}  & \colhead{Median} &  \colhead{Geometric} & \colhead{Standard} \\
\colhead{} & \colhead{Size} & \colhead{} & \colhead{Mean (pc)}  & \colhead{Dev (pc)}  &  \colhead{Error$^{a}$ (pc)}  & \colhead{(pc)} & \colhead{Mean (pc)} & \colhead{Dev Factor$^{a}$} \\
}
\startdata
Milkyway	& 214 & Table \ref{tab:1}         & 30.5 & 25.4	& 1.7 &	24.3 & 21.9	& 2.4	\\
M31			& 156 & \cite{2014LeeLee}        & 48.3 & 19.0	& 1.5 &	46.2 & 44.2	& 1.6	\\
M33			& 137 & \cite{2010Long}           & 49.6 & 28.8	& 2.5 &	44.0 & 42.6	& 1.8	\\
M83			&  47 & \cite{2010Dopita}         & 22.7 & 10.3	& 1.5 &	23.9 & 20.5	& 1.6	\\
LMC			&  59 & \cite{2017Bozzetto}       & 39.0 & 21.3	& 2.8 &	33.8 & 31.6	& 2.3	\\
SMC			&  21 & \cite{2005FilipovicPayne} & 40.8 & 18.4	& 4.0 &	36.0 & 37.5	& 1.5	\\
\enddata
\tablecomments{a - Definitions are given in Section \ref{sizedist}.
 }
\end{deluxetable*}

\indent The SNR samples of most of the galaxies in the above studies have a striking similarity, which is the lack of SNRs with small physical diameters. Only a few SNRs ($< \sim5$) in each sample have SNRs where the diameter is $< 10$ pc. Past authors \citep{2010Long, 2010Badenes} have stated that this deficit is likely a real one. There are $\sim35$ SNRs with diameters $< 10$ pc in our sample. While the relatively large number of SNRs with small physical diameters are present in the Galaxy, the sample of SNRs/candidates presented by \cite{2021Winkler} (and references therein) for M83 has almost a $100$ SNRs with small physical diameters that are likely expanding in locally dense regions of the ISM. Most of the SNRs in other galaxies have been identified using H$\alpha$ and [S II] lines, where the identification of SNRs with smaller diameters is not expected  \citep{2021Winkler}. Galactic SNRs are generally identified using different criteria (see \cite{2006Brogan}) and are observed at higher spatial resolution, which explains the higher number of Galactic SNRs with smaller diameters. \\
\indent The cumulative distribution function (CDF) of Galactic SNR diameters is shown in Figure \ref{fig:3} (right panel). For comparison, the CDFs of five other galaxies (M31, M33, M83, LMC and SMC) are given as well. The Kolmogorov-Smirnov (KS) test was performed on the diameter distributions and showed that the Galactic SNR diameter distribution is not consistent with any of the other distributions. The discrepancies between the diameter distributions could be real but are affected by the different methods of SNR identification.\\

\subsection{Supernova Remnant Distances and $\Sigma$-D Relation}

\indent While kinematic distances to many SNRs have been estimated using HI data, MC or maser associations, optical extinction, etc., there are a considerable number of SNRs with no distances ($\sim1/3$ of the  \cite{2019GreenCat} catalogue and $\sim1/2$ of the \cite{2012Ferrand} catalogue). When there are no other means to estimate the distances to SNRs, the surface brightness ($\Sigma $)- physical diameter (D) relation  ($\Sigma$-D) has been employed. The main idea is that the surface brightness of a SNR changes as the SNR expands or as the diameter increases with time. The $\Sigma $-D relationship between the radio surface brightness at a frequency $\nu$ and the diameter of a SNR is given by 

\begin{equation}
\Sigma_\nu = AD^{-\beta}.
\label{eq:1}
\end{equation} 

With known SNR distances (from other methods) as calibrators, an empirical $\Sigma-D$ relation is determined to estimate the unknown SNR distances. It should be noted that this relation is mainly applicable for shell-type SNRs. However, the majority of the Galactic SNRs are shell-type SNRs ($\sim80\%$ of the Green's SNRs). 

\begin{figure*}
\centering
\includegraphics[width=\textwidth]{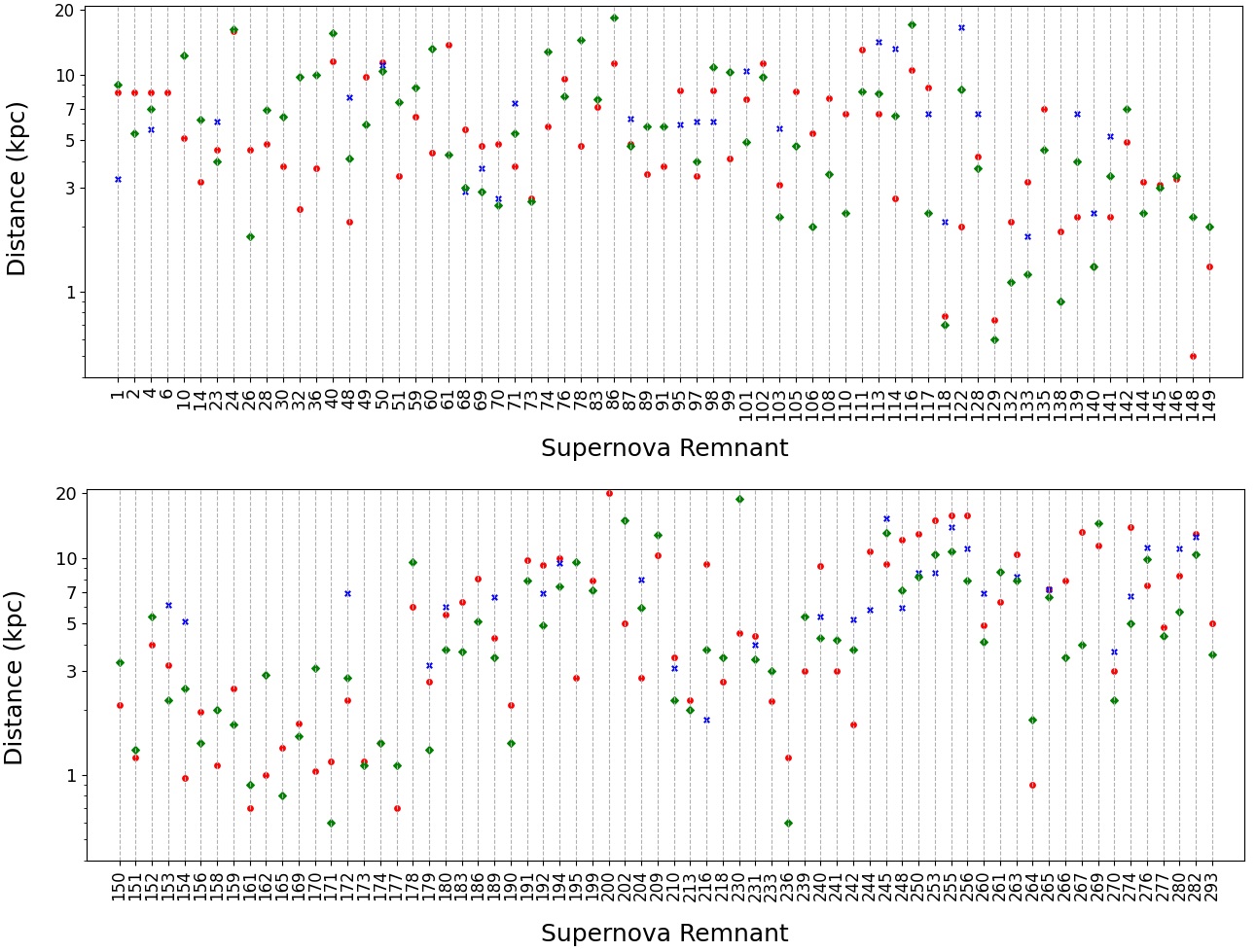}
\caption{A comparison between the literature distances and distances obtained using the $\Sigma$-D relation. The red circles denote the literature distances to SNRs from \cite{2022RanasingheLeahy}. The distances obtained from the $\Sigma$-D relation presented by \cite{1998CaseBhat} and \cite{2014Pavlovic} are denoted by blue `x's and green diamonds, respectively. The numbers in the x-axis labels corresponds to SNR numbers from Table \ref{tab:1} } 
\label{fig:1}
\end{figure*}

\indent Many authors have used the $\Sigma$-D relation in the past to estimate distances to SNRs. Both  \cite{1998CaseBhat}  \& \cite{2014Pavlovic} have presented power-law index values ($\beta = 2.4$ and $\sim5.2 $, respectively) and new distances. Their calibrator sample sizes were 36 and 65 SNRs, respectively.

\begin{figure} [!htb]
\centering
\includegraphics[width=\columnwidth]{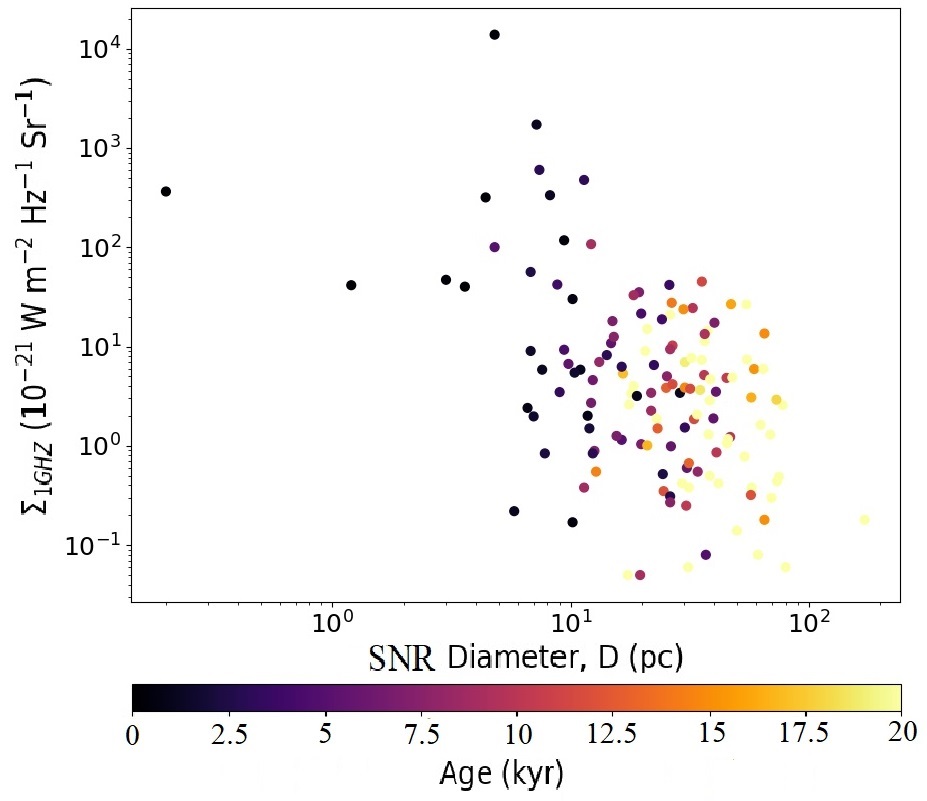}
\caption {Distribution of Galactic SNRs in the  $\Sigma$-D plane for a sample of 139 shell-type SNRs. The colour bar denotes the ages of the SNRs in kyr. }
\label{fig:4}
\end{figure}

\indent Out of the 214 SNRs with diameter estimations, 29 do not have 1-GHz flux densities. Out of the  remaining 187 SNRs 139 SNRs are shell-type. For our sample, we have included the uncertain shell-type SNRs as well. Figure \ref{fig:4} shows the distribution of the SNRs in the $\Sigma$-D plane and there is a clear decrease of $\Sigma$ with increasing $D$, on average. However, there is a large scatter in the distribution and for some SNR diameters there is a wide range of surface brightnesses for different SNRs. For an example, at a  $\sim 80$ pc diameter, the SNR surface brightness has a range of values spanning 3 orders of magnitudes. This scatter in the $\Sigma$-D plane  has been pointed out by \cite{1991Green} and as recently by \cite{2014Pavlovic}. Not all SNRs evolve the same and is affected by parameters such as explosion energy, ambient density, etc. Therefore, a large scatters in the $\Sigma$-D plane are expected \citep{2018Pavlovic}.  Furthermore, the importance of ISM density for the strength of radio luminosity of SNRs was found by \cite{2022LeahyMerrick}, which is not predicted in analytic models for radio emission. While distance estimations using a $\Sigma$-D might be more accurate in other galaxies (e.g. \citealt{2017Bozzetto}), it is unlikely a single relation could be valid for our Galaxy. 

\begin{figure*} [!htb]
\centering
\includegraphics[width=\textwidth]{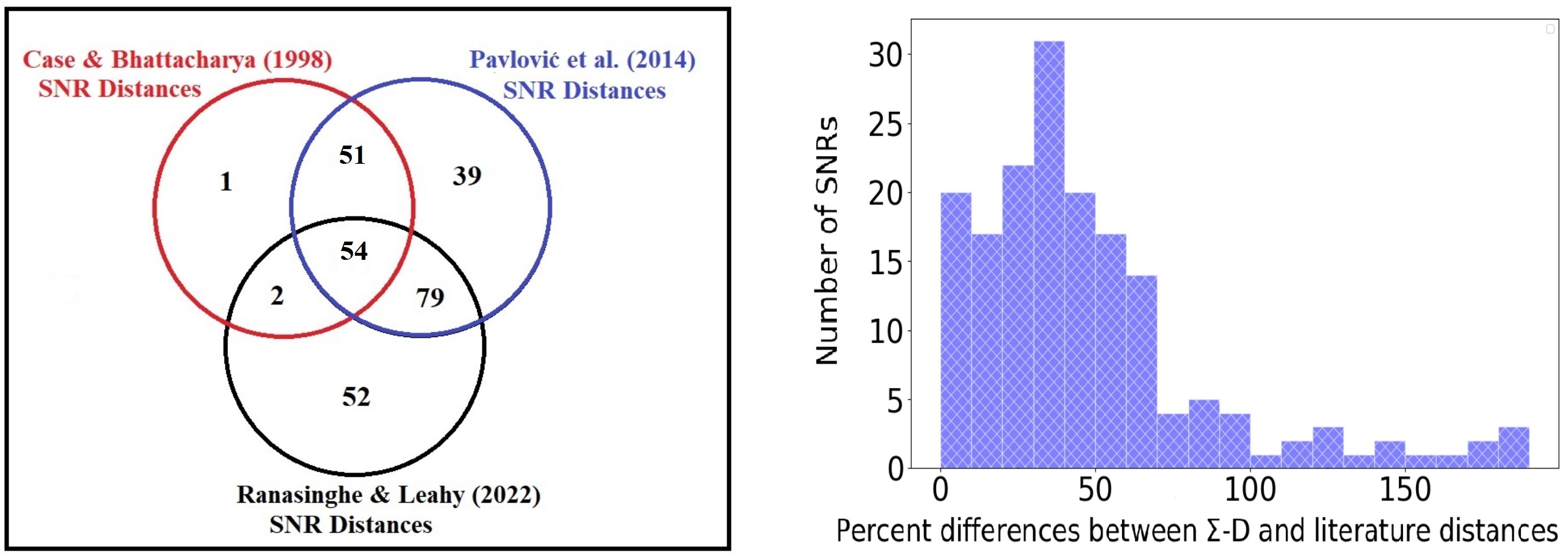}
\caption {Left panel: Venn diagram showing the number of SNRs  with distances \citep{2022RanasingheLeahy} and the number of SNRs with distances estimated using the $\Sigma$-D relation (\citealt{1998CaseBhat} \& \citealt{2014Pavlovic}). Right panel: Histogram of the percentage differences between the literature distances and the $\Sigma$-D distances. }
\label{fig:2}
\end{figure*}

\indent As \cite{1991Green} stated, it may be possible to improve the  $\Sigma$-D relation if the sample was taken from a particular class of SNRs (e.g. SNRs associated with MC). We have investigated the relation taking samples by MC associations, SN type and Galactic height (z) and found none of the $\Sigma$-D distributions provided a better relation. The power law index, $\beta$ from Equation \ref{eq:1} as calculated in SNR radio emission models, explicitly depends on the spectral index, $\alpha$ \citep{2014Pavlovic}. We investigated the $\Sigma$-D relation for different subsets of spectral indices and found that the the different sets of fixed spectral index ranges did not show any reduced scatter with respect to that for the whole set. This may be due to the fact that the spectral indices are an average across the face of SNRs. Regardless, the samples for each spectral index ($\pm0.02$) are too small ($\sim10$ or less) to make a definite conclusion. The  $\Sigma$-D distribution in Figure \ref{fig:4} shows a clear downward trend but only for young SNRs. The older SNRs that are $>20$ kyr in age (diameters between 100 \& 200 pc) are scattered showing no discernible trend.  The ages of SNRs come from literature and estimations from this work (see Section \ref{Ages} for details)  \\
\indent The $\Sigma$-D relation has been used as a first guess when there are no other methods of distance estimations are available. However, the $\Sigma$-D distances could differ from the actual distance by an order of a magnitude and the large scatter in the Galactic $\Sigma$-D distribution leads to large errors. Figure \ref{fig:1} shows the literature distances from \cite{2022RanasingheLeahy} and  $\Sigma$-D distances (\citealt{1998CaseBhat} \& \citealt{2014Pavlovic}). As seen in Figure \ref{fig:1}, only a few $\Sigma$-D distances agree with the literature values. For this comparison we only use the \cite{2019GreenCat} SNRs (sample size: 294). Figure \ref{fig:2} left panel shows a Venn diagram with the number of SNRs in each study. The right panel of Figure \ref{fig:2} shows the percent difference between the  $\Sigma$-D and literature distances. There are only 20 distances that are  within $5\%$ of each other. The majority differ at least by $20\%$ (on average differ by $50\%$). Thus, while $\Sigma$-D distances may give an idea of SNRs' location, the reliability of the result is questionable.     
 
\subsection{Supernova Remnant Ages} \label{Ages}
 
\indent The five main methods of estimating SNR ages use i) age of associated pulsar ii) the explosion energy of the progenitor and the energy loss rate of the SNR, iii) the expansion velocity and radius of the SNR, iv) the thermal temperature measured from X-ray observations, and v) the remnant's measured spectral break due to synchrotron losses and its magnetic field \citep{2005Xu}. However, the uncertainties of the ages are large (up to an order of magnitude) and the parameters needed to estimate the ages for many SNRs are unknown. \\
\indent Out of the 390 SNRs in our sample, there are 141 SNRs with literature ages, where 9 are given as a lower or upper limit. Of the 141 SNR, Sedov ages to 22 SNRs were presented with assumed parameters (E$_0$ and n$_0$). The remaining 119 literature ages were estimated using more accurate methods (e.g. X-ray data, shock velocity).\\
\indent We estimate ages for 75 SNRs with no previously published ages and present revised ages to the 22 SNRs with previously published Sedov ages (97 SNR ages in total). To estimate the age, we assume that the shock radius of each SNR to be the average radio radius. With no additional information (e.g. distance, radius), 174 SNRs in the sample have no ages.    \\
\indent  To estimate the ages we use the software presented by \cite{2017LeahyWilliams}. Here we assume the  explosion energy is assumed as, E$_0$ $= 0.75 \times 10^{51}$ erg, a value between the standard explosion energy of E$_0$ = $10^{51}$ erg and the  logarithmic mean explosion energy of $ 0.5 \times 10^{51}$ erg presented by \cite{2017Leahy} (for LMC SNRs). If the SN explosion type of the resulting remnant is known, the ejected mass was assumed to be 1.4M$_{\sun}$ for type Ia SN event and  5M$_{\sun}$ for core-collapse (CC). The majority (about 85\%; \citealt{1994Tammann}) of the SNe are CC, where the progenitor is a young massive star. Therefore, if the progenitor type of a SNR is unknown, we assume the ejected mass to be 5M$_{\sun}$. Furthermore, we assume the interstellar (ISM) number density, $n_0$ to be 10 cm$^{-3}$ if the SNR is associated with MC and 0.01 cm$^{-3}$ if it is not. For probable or unknown MC associations, we take the  $n_0$ to be 1 cm$^{-3}$. However, when available, we use the literature number density values. The ages were estimated for an ejecta power-law index, n = 7 and the circumstellar medium (CSM) power-law index, s = 0.\\
\indent The assumed explosion energies and ISM densities could differ by a few orders of magnitudes from the actual values. \cite{2018LeahyRanasinghe} and \cite{2020LeahyRanasingheGelowitz} presented ages for 58 SNRs, estimated using X-ray shock temperatures and emission measures and provided explosion energies and ISM densities. As a test to evaluate the accuracy of SNR age estimations described above, we re-calculated the ages for these 58 SNRs using the fixed assumed explosion energies and densities (E$_0$ $= 0.75 \times 10^{51}$ erg and $n_0$ to be 1 cm$^{-3}$ ). Most of the ages differ by factors of $\sim1/4$ to $\sim4$ times (on average 1) compared to the ages estimated using X-ray data. In some cases, the SNR ages differed by an order of magnitude. However, considering   that majority of the SNR ages fall between  $\pm 50\% $, we adopt it as the uncertainty of SNR ages estimated in this work. \\ 
\indent It should be noted that some of the ages estimated using the above method are likely incorrect because of incorrect assumptions. For an example, G$65.8-0.5$ at a distance of $2.4^{+0.3}_{-0.5}$ kpc gives an age of 500 yr for an E$_0$ $= 0.75 \times 10^{51}$ erg and $n_0 = 1$ cm$^{-3}$. The age for G$65.8-0.5$ is likely $>500$ yr with either a lower explosion explosion energy or higher ISM density. If the explosion energy is lower by 2 orders of magnitude or the ISM density is higher by 2 orders of magnitude, the age of G$65.8-0.5$ is 6000 and 3500 yr, respectively.  However, without additional information, it is difficult to constrain the ages for SNRs such as G$65.8-0.5$. Therefore, we do not discard these ages from further analysis.  
 
\begin{figure} [!htb]
\centering
\includegraphics[width=\columnwidth]{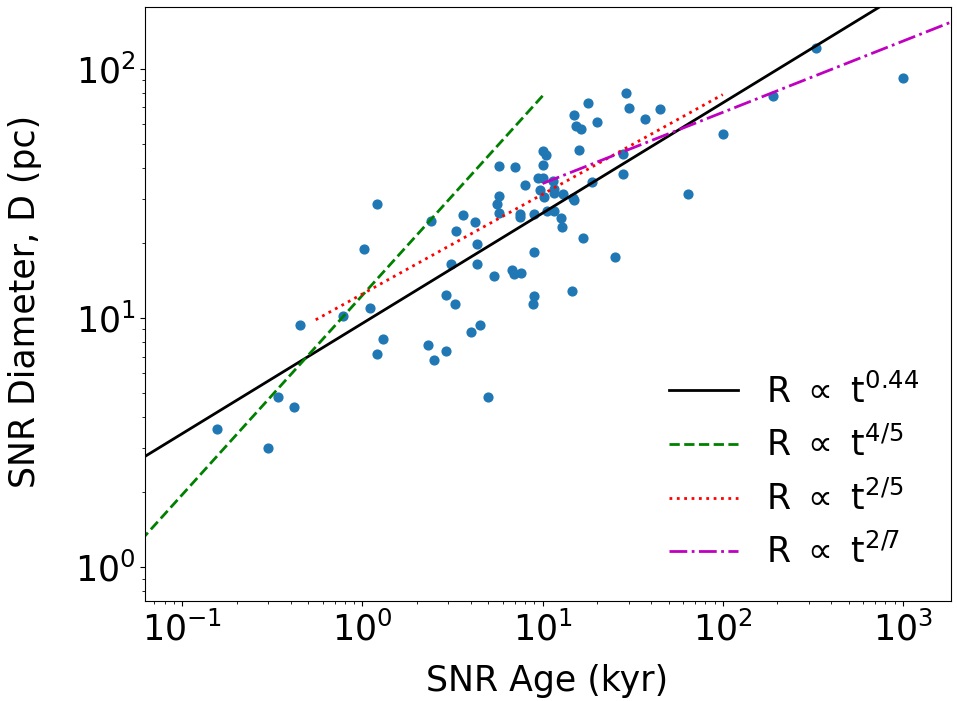}
\caption {Galactic SNR diameters and their age distribution for a sample of 80 shell-type SNRs. The best fit line (black curve) is  $ D = (9.52 \pm 0.92)$ $t^{0.44 \pm 0.04} $ pc. The green dashed line is $r \propto t^{4/5}$ ejecta-dominated stage, the red dotted line is $r \propto t^{2/5}$ adiabatic stage magenta dash-dotted line $r \propto t^{2/7}$ radiative stage.  }
\label{fig:5}
\end{figure} 

\indent Figure \ref{fig:5} shows the plot of SNR diameter against their age. To determine the diameter-age ($D-t$) relation we have only included the literature ages estimated using more reliable methods (e.g Pulsar associations, expanding velocity, X-ray observation, etc.). The sample consists of 83 shell-type SNRs that have literature ages. For comparison, in Figure \ref{fig:5} we have included lines for the main stages of SNR evolution. The ejecta-dominated stage (green dashed line) has $r \propto t^{4/5}$ \citep{1982Chevalier}, adiabatic stage (red dotted line) has $r \propto t^{2/5}$ \citep{1959Sedov} and the radiative stage (magenta dash-dotted line) has $r \propto t^{2/7}$ \citep{1988Cioffi}. Here $r$ is the shock radius and $t$ is the age of the SNR.  The best-fit line for the data is given by 

\begin{equation}
\left( \frac{D}{\textrm{pc}} \right) = (9.52 \pm 0.92) \left( \frac{t}{\textrm{kyr}} \right)^{0.44 \pm 0.04}.
\label{eq:2}
\end{equation} 

\indent Similar to the $\Sigma$-D distribution, the $\Sigma$-t (surface brightness - age) distribution shows a clear downward trend. However, a single relation does not fit the data because of the scatter spanning 2-3 orders of magnitudes. Nevertheless, the downward trend does confirm that the SNR surface brightness diminishes with age on average.     

\begin{figure*} [!htb]
\centering
\includegraphics[width=\textwidth]{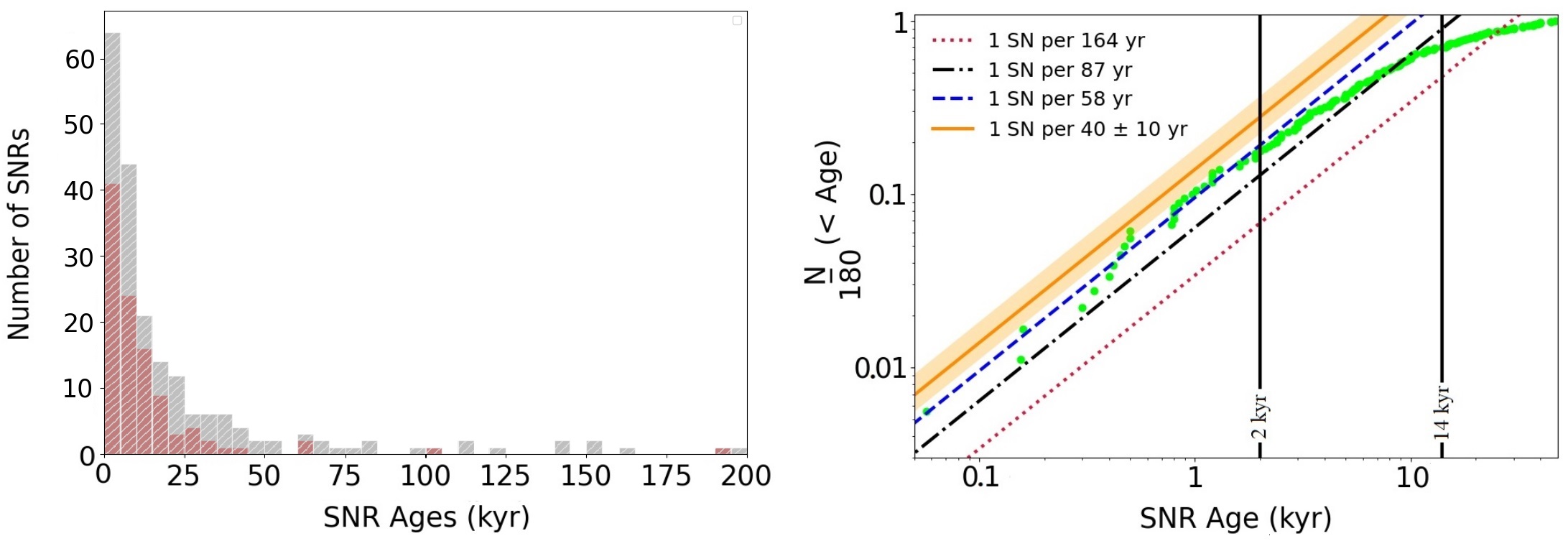}
\caption {Left: Histogram of literature SNR ages (red) and all SNR ages in our sample (grey). Right: The cumulative distribution of SNR Ages (180 youngest sources). The best fit lines: 1 SN in 164 yr (for ages $<50$ kyr, red dotted line, sample size: 180), 1 SN in 87 yr (for ages $<14$ kyr, black dash-dot line, sample size: 125) and 1 SN in 58 yr (for ages $<2$ kyr, blue dashed line, sample size: 31). The orange line is the SN rate of 1 SN in $40 \pm 10$ yr presented by \cite{1994Tammann} with uncertainty shown by the orange shaded region. }
\label{fig:8}
\end{figure*}

\indent Figure \ref{fig:8} left panel shows the histogram of the literature SNR ages in red and all SNR ages (includes literature ages and estimations from this work) in our sample in grey. Most of the literature ages ($> 95\%$) are $< 65$ kyr. The main reason for the lack of older SNRs  is that they are not bright enough to have an easily measured X-ray spectrum \citep{2018LeahyRanasinghe}. The sparsity of older SNRs ($> \sim50 $) is apparent in both samples (literature ages and all) reiterating the incompleteness of the sample of SNRs with ages. 
\indent Figure \ref{fig:8} right panel shows the cumulative age distribution of Galactic SNRs (youngest 180 in ascending order). There are 204 SNRs with ages in the sample (excluding ages given as upper and lower limits). The best-fit line for the youngest 180 SNR ages gives a SN rate of 1 SN per 164 yr with a correlation coefficient of $\sim0.87$. Owing to the incompleteness of the sample of SNRs with known ages, the best-fit line of 1 SN per 164 yr does not describe the trend well (Figure \ref{fig:8} red dotted line compared to the data in green).\\
\indent The youngest 125 ($< 14$ kyr) and 31 ($< 2$ kyr) SNR ages in the sample give better best-fit lines of 1 SN per 87 yr and 1 SN per 58 yr,  respectively,  both with correlation coefficients of $>0.98$. While the best-fit line of 1 SN per 58 yr is smaller than the lower limit of the SN rate presented by \cite{1994Tammann}, it is within $2\sigma$. Furthermore, 1 SN per 58 yr rate is consistent with the CC SN rate of 1 per $61^{+24}_{-14}$ yr presented by \cite{2021Rozwadowska}. From Figure \ref{fig:8} it is seen that the incompleteness of the sample for older SNRs increases.  The sample is likely nearly complete for the SNR ages $< 2$ kyr, but the incompleteness is a factor of $\sim2$ for SNR ages $< 14$ kyr.\\

\subsection{Radio Spectral Indices of Supernova Remnants}

\begin{figure*} 
\centering
\includegraphics[width=\textwidth]{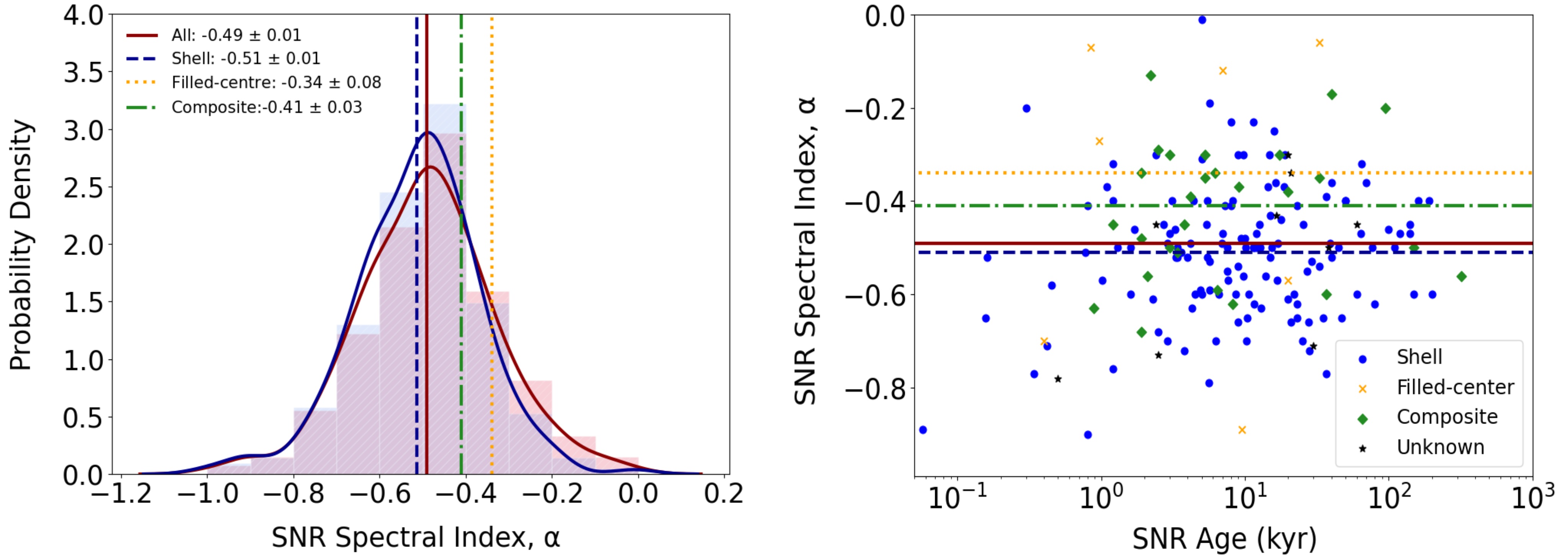}
\caption {Left panel: The PDF of the spectral index ($\alpha$) distribution. SNRs (all types: 270 SNRs in sample) in red (histogram and PDF) and shell-type SNRs (208 sample size) in blue (histogram and PDF). The vertical lines show the mean spectral index for each SNR type. Right panel: The spectral index vs age (175 SNRs in sample). The horizontal lines show the mean spectral index for each SNR type (same as left panel). }
\label{fig:6}
\end{figure*}
 
A non-thermal spectral index ( $\alpha < 0$) is one the criteria in identifying a SNR \citep{2006Brogan}.  Figure \ref{fig:6} left panel shows the PDF of the Galactic SNR spectral index distribution. We find the mean spectral index of all Galactic SNRs to be $ \alpha = -0.49 \pm 0.01 $ with a $0.16$ Standard deviation. The estimated error is the standard error. The statistics on spectral indices of SNRs  are given in Table \ref{tab:3}. The mean spectral index of the shell-type SNRs is $ \alpha = -0.51 \pm 0.01 $. Both mean spectral indices are consistent with the theoretical SNR spectral index of $-0.5$. The mean spectral index of the LMC SNRs is $\alpha = -0.52 $ \citep{2017Bozzetto} and of the SMC SNRs is  $\alpha = -0.63 $ \citep{2005FilipovicPayne}. We find the mean Galactic SNR (all and shell-type) spectral indices are consistent with the LMC mean index, however, the SMC SNR/candidate mean spectral index is steeper than the Galactic mean index.  \\
\indent The mean composite SNR spectral index is $ \alpha = -0.41 \pm 0.03 $, which is slightly flatter than the mean shell-type SNR spectral index. There are 37 composite SNRs in the sample including 17 uncertain composite type SNRs (denoted with a `?'). The composite spectral indices range from $-0.68$ to $-0.1 $ with a standard deviation of $0.15$.  However, it should be noted that the indices $< -0.2 $ are classified as uncertain composite SNRs. 

\begin{deluxetable}{ccccc}
\tablenum{3}
\label{tab:3}
\tablecaption{Statistics on SNR radio spectral indices}
\tablewidth{700pt}
\tabletypesize{\small}
\tablehead{
\colhead{SNR}  & \colhead{Mean} & \colhead{Standard} & \colhead{Standard } & \colhead{Median}  \\
\colhead{type} & \colhead{}     & \colhead{dev}      & \colhead{error}    & \colhead{} \\
}
\startdata
All	            & -0.49 &  0.16 & 0.01  & -0.50 \\
Shell			& -0.51 &  0.14 & 0.01  & -0.50	\\
Filled-center   & -0.34 &  0.27 & 0.08  & -0.26	\\
Composite		& -0.41 &  0.15 & 0.03  & -0.39 \\
Unknown         & -0.49 &  0.17 & 0.05  & -0.45 \\  
\enddata
\end{deluxetable}

\indent The mean filled-center SNR spectral index is $ \alpha = -0.34 \pm 0.08 $. The spectral index of filled-center SNRs are  characterized as being flat where $\alpha \sim 0 - -0.3$ \citep{1978Weiler}. There are 11 filled-center SNRs and majority of the spectral indices fall in the range of $\alpha \sim 0 - -0.3$. In fact only 3 filled-center SNRs have steeper spectral indices (G$65.7 +1.2$, G$76.9+1.0$ and G$141.2+5.0$). If the three steeper spectral indices are excluded from the filled-center SNR sample, the mean spectral index is $-0.20 \pm 0.03$ with a standard deviation of $0.10$. While the filled-center SNR sample size is too small to form a definite conclusion, both mean spectral indices (total sample and sample excluding the three steeper $\alpha$) are consistent with the generally observed filled-center SNR spectral index. \\  
\indent Additionally, we have analysed the spectral index distribution of known pulsar wind nebulae (PWNe). The PWNe with radio counterparts were taken from the catalogue presented by \cite{2004RobertsPWN}\footnote{\url{https://www.physics.mcgill.ca/~pulsar/pwncat.html}}. Similar to filled-center SNRs, the spectral indices of PWNe are expected to be relatively flat. There are 34 PWNe in the \cite{2004RobertsPWN} catalogue that have associated SNRs, where 31 have spectral indices (3 have varying $\alpha$s across the SNRs). The PWNe spectral indices range from $-0.68 $ to $-0.06 $. The mean spectral index of the PWN sample is $-0.36 \pm 0.03$ with a standard deviation of $0.17$. However, unlike the filled-center SNR spectral index distribution, the PWNe spectral indices appear to follow a normal distribution.  \\  
\indent We have investigated the spectral index distribution against MC associations, SN type, diameter  and surface brightness and found no correlation between them confirming the conclusion formed by \cite{1976ClarkCaswell}. The age-spectral index plot (Figure \ref{fig:6}, right panel) shows significant scatter and no apparent trend. Younger SNRs are expected to have a flatter spectral index \citep{2015Dubner}, but the majority of the young SNRs ($< 1$ kyr) in our sample are seen to have a steeper spectral index (average of $\sim0.7$). The older SNRs ($> 100$ kyr) show an average spectral index close to the theoretical value of $\sim0.5 $. This correlation, while weak, seems to be consistent with the LMC SNR spectral index distribution (see \citealt{2017Bozzetto}).

\subsection{Galactic Height Distribution of SNRs}

\begin{figure} [!htb]
\centering
\includegraphics[width=\columnwidth]{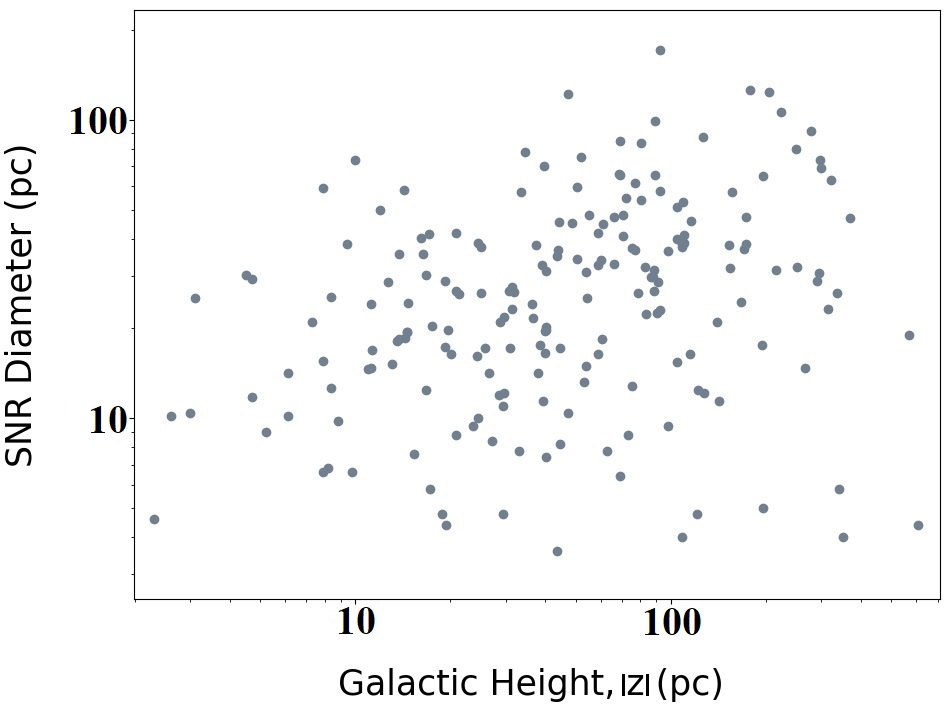}
\caption {The Galactic radio SNR diameters, D vs. Galactic height, z for the sample of 214 radio SNRs.}
\label{fig:9}
\end{figure}

\indent The majority of the SNRs are located in the Galactic plane and most of them are CC that can be associated with star forming regions. We have investigated whether there is a correlation between the SNR diameter and the Galactic height, z. There is a large scatter in the distribution and the correlation of size with Galactic height is poor (Figure \ref{fig:9}). However, while the correlation is poor, there is an upward trend, indicating that as the galactic height increases the SNR diameter increases. Furthermore, the scatter in the $D-z$ plane is not further reduced when considering only SNRs with or without MC associations.

\begin{figure} [!htb]
\centering
\includegraphics[width=\columnwidth]{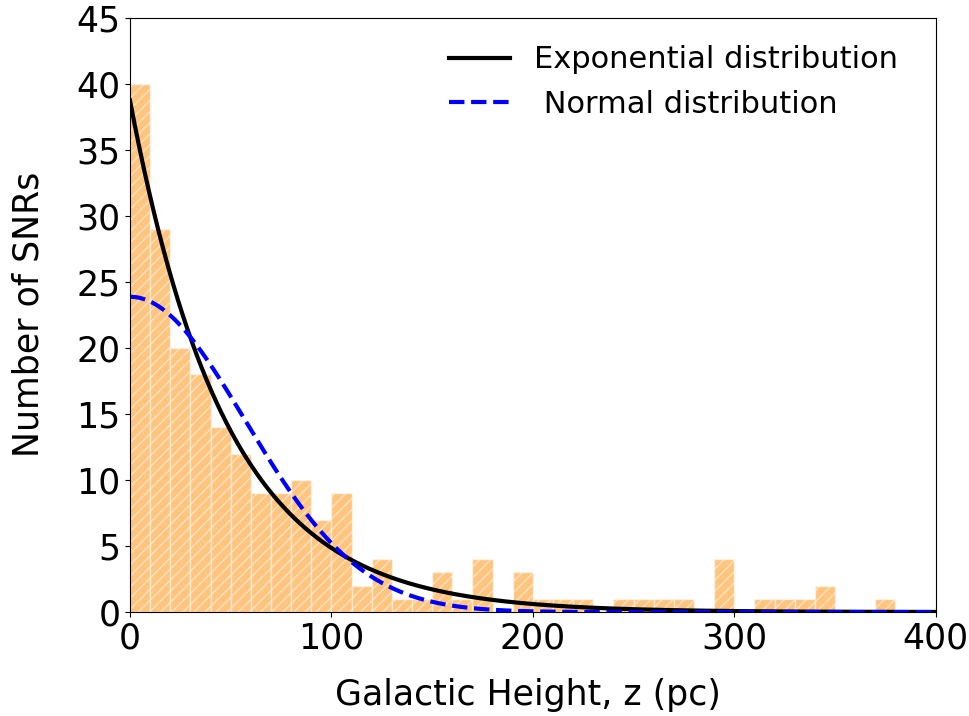}
\caption {The Galactic height distribution (z) distribution of SNRs (sample size of 214). The best-fit exponential distribution, $N = 38.8$ exp($ -\lvert z \rvert/48.1$) is the black solid curve. The blue curve is the best-fit normal distribution, $N = 23.9$ exp($ -0.5(x/57.3)^2$).  }
\label{fig:7}
\end{figure}

\indent Figure \ref{fig:7} shows the Galactic height distribution of SNRs. We fit exponential and normal distributions to the data (sample size of 214). The best-fit line for the exponential distribution is given by $ N = 38.8 $ exp($ -\lvert z \rvert /48.1 $) and for the normal distribution is given by $N = 23.9$ exp($ -0.5(x/57.3)^2$). The minimized $\chi ^2$ values for the 2 models are 28.2 and 41.5, respectively. With a  degree of freedom of 59, the p-value for the exponential distribution is $>0.99$ and for the normal distribution is 0.96. Therefore, examining the curves and p values, we conclude that the exponential distribution describes the data best.\\
\indent For the Galactic height exponential distribution of SNRs, most of the SNRs are located near the galactic plane. In fact $ \sim 90\% $ of the SNRs are located at a Galactic height $ < 200 $ pc. This is consistent with the massive star and star forming region distribution of the Galaxy \citep{2014Urquhart, 2016BlandHawthorn}.\\

\subsection{Spherical Symmetry of SNRs}
\indent The spherical symmetry of SNRs has been discussed by many authors in the past. \cite{2009Lopez} in their study from X-ray data presented evidence that SNRs can be be separated by SNe type (Ia or CC). Similarly, the infrared (IR) study done by \cite{2013Peters} shows that the SNRs resulting of a type Ia SN have  a more circular and mirror symmetric morphology. However, as stated by \citep{2015Dubner} and demonstrated by \cite{2019Ranasinghe}, there is no discernible  separation of SNR types according to their radio morphology. This may be caused by the complex interactions between the SNRs and the ISM. \\
\indent We have examined whether there is a correlation between the ovality and other SNR parameters. Here the ovality is estimated as defined by  \cite{2017Bozzetto} (their equation 9). We found that none of the SNR parameters (spectral indices, SNR ages, MC associations and galactic heights) or the explosion type was correlated to the spherical symmetry of SNRs.

\section{Conclusions} \label{conclusions}

We have compiled a table of 390 Galactic radio SNRs and their basic parameters. The following conclusions are formed from our analyses.

\begin{enumerate}

\item{The arithmetic mean of the Galactic SNR diameters is $30.5$ pc with standard error $1.7$ pc and standard deviation $25.4$ pc. The geometric mean and geometric standard deviation factor of the Galactic radio SNR diameters is $21.9$ pc and $2.4$, respectively. While the arithmetic mean of Galactic SNR diameters is different from the mean SNR diameters from other galaxies, the geometric mean of Galactic SNR diameters is comparable to the SNR diameter means of M83.}
\item{The $\Sigma$-D relation for the Galactic SNRs shows a large scatter in the $\Sigma$-D plane because the radio brightness is affected by a large number of physical factors \citep{2018Pavlovic}. We did not find the scatter to reduce for particular classes of SNRs. A comparison of distances obtained by different $\Sigma$-D relations to distances obtained by more reliable methods show a discrepancy of $50\%$ on average, questioning the reliability of the  $\Sigma$-D method. }
\item{We have estimated ages to 97 SNRs and on average $ D = (9.52 \pm 0.92)$ $t^{0.44 \pm 0.04} $ pc for shell-type SNRs. However, this relation has a large scatter of about factor 2 (Figure \ref{fig:5}).}
\item{A birthrate of 1 SN in 58 yr fits the SNR age distribution for $t< 2$ kyr and is consistent with the currently accepted SN birth rate within $2 \sigma $. As age (t) increases, the mean birthrate drops, indicating the incompleteness of the sample of SNRs increases with age. }
\item{The mean spectral index of the Galactic shell-type SNRs is $-0.51 \pm 0.1$. No correlations were found between the spectral index values and other SNR parameters (i.e. MC associations, SN type, diameter  and surface brightness). On average, the younger SNRs have a steeper spectral index.}
\item{The majority of the SNRs are located near the Galactic plane and the Galactic height distribution of SNRs is best described by an exponential with scale height $48 \pm 4$ pc. }
\item{The spherical symmetry of a SNRs in radio is not correlated to any other SNR parameter or the explosion type. }

\end{enumerate}

We thank the referee for the comments and suggestions that have improved this work. \\
We acknowledge the support of the Natural Sciences and Engineering Research Council of Canada (NSERC).\\
Nous remercions le Conseil de recherches en sciences naturelles et en génie du Canada (CRSNG) de son soutien.

\begin{turnpage}
\begin{deluxetable*}{llccccccccccccc}
\tablenum{1}
\label{tab:1}
\tablecaption{Galactic SNRs and their parameters}
\tablewidth{700pt}
\tabletypesize{\scriptsize}
\tablehead{
\colhead{\#} & \colhead{SNR}  & \colhead{Size} &  \colhead{SNR}  & \colhead{1-GHz } &  \colhead{Spectral } & \colhead{Heliocentric} & \colhead{Galactocentric} & \colhead{Height, z} & \colhead{MC} & \colhead{Age$^b$} & \colhead{SN} & \colhead{Avgerage} & \colhead{Surface} & \colhead{Ovality}\\
\colhead{} & \colhead{ }  & \colhead{}  &  \colhead{type$^a$}  & \colhead{Flux density} & \colhead{index, $\alpha $} & \colhead{distance} & \colhead{distance} & \colhead{} & \colhead{interact} & \colhead{} & \colhead{Type} & \colhead{radius$^c$} & \colhead{Brightness $^d$} & \colhead{\%}\\
\colhead{} & \colhead{ }  & \colhead{(arcmin)}  &  \colhead{}  & \colhead{(Jy)} & \colhead{} & \colhead{(kpc)} & \colhead{(kpc)} & \colhead{(pc)} & \colhead{} & \colhead{(kyr)} & \colhead{} & \colhead{(pc)} & \colhead{} & \colhead{}\\
}
\startdata
1	&  G$0.0+0.0	$	&$	3.5	\times	2.5	$&	S	&	100?	&$	-0.76			$&$	8.34	\pm	0.16	$&$	0			$&$	0	$&	Yes	&$	1.2		$&	Iax	&$	3.6	\pm	0.1	$&$	1719.69?	$&	33	\\
2	&  G$0.3+0.0	$	&$	15	\times	8	$&	S	&	22	&$	-0.56	\pm	0.1	$&$	8.34	\pm	0.16	$&$	0.04			$&$	0	$&	\nodata	&$	14*		$&	\nodata	&$	13.9	\pm	0.3	$&$	27.59	$&	61	\\
3	&  G$0.9+0.1	$	&$	8	\times	8	$&	C	&	18?	&$	-0.68	\pm	0.07	$&$	8.34	\pm	0.16	$&$	0.13			$&$	+14.6	\pm	0.3	$&	Yes 	&$	1.9		$&	\nodata	&$	9.7	\pm	0.2	$&$	42.32?	$&	0	\\
4	&  G$1.0-0.1	$	&$	8	\times	8	$&	S	&	15	&$	-0.6?			$&$	8.34	\pm	0.16	$&$	0.15			$&$	-14.6	\pm	0.3	$&	\nodata	&$	7*		$&	\nodata	&$	9.7	\pm	0.2	$&$	35.27	$&	0	\\
5	&  G$1.4	-0.1	$	&$	10	\times	10	$&	S	&	2?	&	\nodata			&	$8.5 $	 		&	\nodata			&	\nodata			&	Yes	&	\nodata		&	\nodata	&	\nodata			& $	3.01?	$&	0	\\
6	&  G	$	1.9+0.3	$	&$	1.5	\times	1.5	$&	S	&	0.6	&$	-0.65			$&$	8.34	\pm	0.16	$&$	0.28			$&$	+43.7	\pm	0.8	$&	Yes	&$	0.156	\pm 0.011	$&	Ia	&$	1.8	\pm	0.0	$&$	40.13	$&	0	\\
7	&  G$3.7-0.2	$	&$	14	\times	11	$&	S	&	2.3	&$	-0.65	\pm	0.05	$&	\nodata			&	\nodata			&	\nodata			&	\nodata	&	\nodata		&	\nodata	&	\nodata			&$	2.25	$&	24	\\
8	&  G$3.8+0.3	$	&$	18	\times	18	$&	S?	&	3?	&$	-0.6	\pm	0.1	$&	\nodata			&	\nodata			&	\nodata			&	\nodata	&	\nodata		&	\nodata	&	\nodata			&$	1.39?	$&	0	\\
9	&  G$4.2-3.5	$	&$	28	\times	28	$&	S	&	3.2?	&$	-0.6			$&	\nodata			&	\nodata			&	\nodata			&	\nodata	&	\nodata		&	\nodata	&	\nodata			&$	0.61?	$&	0	\\
10	&  G$4.5+6.8	$	&$	3	\times	3	$&	S	&	19	&$	-0.71	\pm	0.01	$&$	5.1_{-0.7}^{+0.8}	$&$	3.3_{-0.7}^{+0.8}	$&$	+608.1_{-83.5}^{+95.4}	$&	\nodata	&$	0.418		$&	Ia	&$	2.2_{-0.3}^{+0.4}$&$	317.66	$&	0	\\
11	&  G$4.8+6.2	$	&$	18	\times	18	$&	S	&	3	&$	-0.57	\pm	0.13	$&	\nodata			&	\nodata			&	\nodata			&	\nodata	&	\nodata		&	\nodata	&	\nodata			&$	1.39	$&	0	\\
12	&  G$5.2-2.6	$	&$	18	\times	18	$&	S	&	2.6?	&$	-0.6			$&	\nodata			&	\nodata			&	\nodata			&	\nodata	&	\nodata		&	\nodata	&	\nodata			&$	1.21?	$&	0	\\
13	&  G$5.4-1.2	$	&$	35	\times	35	$&	C?	&	35?	&$	-0.2			$&$	5	\pm	0.3	$&$	3.4	\pm	0.3	$&$	-104.7	\pm	6.3	$&	Yes	&$	95*		$&	CC 	&$	25.5	\pm	1.5	$&$	4.30?	$&	0	\\
14	&  G	$	5.5+0.3	$	&$	15	\times	12	$&	S	&	5.5	&$	-0.7	\pm	0.1	$&$	3	\pm	0.7	$&$	5.3	\pm	0.7	$&$	+15.7	\pm	3.7	$&	Yes	&	6.3*		&	\nodata	&$	5.9			$&$	4.60	$&	22	\\
15	&  G$5.9+3.1	$	&$	20	\times	20	$&	S	&	3.3?	&$	-0.4?			$&	\nodata			&	\nodata			&	\nodata			&	\nodata	&	\nodata		&	\nodata	&	\nodata			&$	1.24?	$&	0	\\
16	&  G$6.1+0.5	$	&$	18	\times	12	$&	S	&	4.5	&$	-0.9			$&	\nodata			&	\nodata			&	\nodata			&	Yes?	&	\nodata		&	\nodata	&	\nodata			&$	3.13	$&	40	\\
17	&  G$6.1+1.2	$	&$	30	\times	26	$&	F	&	4.0?	&$	-0.3			$&	\nodata			&	\nodata			&	\nodata			&	\nodata	&	\nodata		&	\nodata	&	\nodata			&$	0.77?	$&	14	\\
18	&  G$6.4-0.1	$	&$	48	\times	48	$&	C	&	310	&$	-0.35	\pm	0.18	$&$	1.8	\pm	0.3	$&$	6.6	\pm	0.3	$&$	-3.1	\pm	0.5	$&	Yes	&$	33		$&	CC	&$	12.6	\pm	2.1	$&$	20.25	$&	0	\\
19	&  G$6.4+4.0	$	&$	31	\times	31	$&	S	&	1.3?	&$	-0.4			$&	\nodata			&	\nodata			&	\nodata			&	\nodata	&	\nodata		&	\nodata	&	\nodata			&$	0.20?	$&	0	\\
20	&  G$6.5-0.4	$	&$	18	\times	18	$&	S	&	27	&$	-0.6	\pm	0.1	$&	\nodata			&	\nodata			&	\nodata			&	Yes?	&	\nodata		&	\nodata	&	\nodata			&$	12.54	$&	0	\\
21	&  G$7.0-0.1	$	&$	15	\times	15	$&	S	&	2.5?	&$	-0.5	\pm	0.15	$&	\nodata			&	\nodata			&	\nodata			&	\nodata	&	\nodata		&	\nodata	&	\nodata			&$	1.67?	$&	0\\
22	&  G$7.2+0.2	$	&$	12	\times	12	$&	S	&	2.8	&$	-0.55	\pm	0.05	$&	\nodata			&	\nodata			&	\nodata			&	Yes?	&	\nodata		&	\nodata	&	\nodata			&$	2.93	$&	0	\\
23	&  G$7.7-3.7	$	&$	22	\times	22	$&	S	&	11	&$	-0.32	\pm	0.05	$&$	4.5	\pm	1.5	$&$	3.9	\pm	1.4	$&$	-291.0	\pm	97.0	$&	\nodata	&$	1.2	\pm 0.6	$&	IIP	&$	14.4	\pm	4.8	$&$	3.42	$&	0	\\
24	&  G$8.3+0.0	$	&$	5	\times	4	$&	S	&	1.2	&$	-0.65	\pm	0.05	$&$	15.9	\pm	1	$&$	7.7	\pm	1.0	$&$	0	$&	Yes	&	47*	&	\nodata	&$	10.4			$&$	9.03	$&	22	\\
25	&  G$8.7-5.0	$	&$	26	\times	26	$&	S	&	4.4	&$	-0.3			$&	\nodata			&	\nodata			&	\nodata			&	\nodata	&	\nodata		&	\nodata	&	\nodata			&$	0.98	$&	0	\\
26	&  G$8.7-0.1	$	&$	45	\times	45	$&	S?	&	80	&$	-0.5			$&$	4.5	\pm	0.3	$&$	3.9	\pm	0.3	$&$	-7.9	\pm	0.5	$&	Yes	&$	15.3 \pm 0.3		$&	\nodata	&$	29.5	\pm	2.0	$&$	5.94	$&	0	\\
27	&  G$8.9+0.4	$	&$	24	\times	24	$&	S	&	9	&$	-0.6			$&	\nodata			&	\nodata			&	\nodata			&	Yes?	&	\nodata		&	\nodata	&	\nodata			&$	2.35	$&	0\\
28	&  G$9.7+0.0	$	&$	15	\times	11	$&	S	&	3.7	&$	-0.6	\pm	0.1	$&$	4.8	\pm	0.2	$&$	3.7	\pm	0.2	$&$	0	$&	Yes	&	22*		&	\nodata	&$	9.1	\pm	0.4	$&$	3.37	$&	31	\\
29	&  G$9.8+0.6	$	&$	12	\times	12	$&	S	&	3.9	&$	-0.45			$&	\nodata			&	\nodata			&	\nodata			&	\nodata	&	\nodata		&	\nodata	&	\nodata			&$	4.08	$&	0	\\
30	&  G$9.9-0.8	$	&$	12	\times	12	$ &	S	&	6.7	& $	-0.4 $ & $	3.8	\pm	0.3	$ & $	4.6	\pm	0.3	$&$	-53.1	\pm	4.2	$ &	Yes	&	8.2*	&	\nodata	& $	6.6	\pm 0.5$ & $ 7.00 $ &	0	\\
31	&	G$10.5 +0.0$	&	$6 \times 6$	&	S	&	0.9	&	$-0.6 \pm 0.1$	&	\nodata	&	\nodata	&	\nodata	&	Yes	&	\nodata	&	\nodata	&	\nodata	&	3.76		&	0	\\
32	&	G$11.0 +0.0$	&	$11 \times 9$	&	S	&	1.3	&	$-0.6 \pm 0.1$	&	$2.4 \pm 0.7$	&	$6.0 \pm 0.7$	&	0	&	Yes	&	1.6*	&	\nodata	&	$3.5 \pm 1.1$	&	1.98		&	20	\\
33	&	G$11.1 -1.0$	&	$18 \times 12$	&	S	&	5.8	&	$-0.41 \pm 0.02$	&	\nodata	&	\nodata	&	\nodata	&	Yes?	&	\nodata	&	\nodata	&	\nodata	&	4.04		&	40	\\
34	&	G$11.1 -0.7$	&	$11 \times 7$	&	S	&	1	&	$-0.75 \pm 0.05$	&	\nodata	&	\nodata	&	\nodata	&	\nodata	&	\nodata	&	\nodata	&	\nodata	&	1.95		&	44	\\
35	&	G$11.1 +0.1$	&	$12 \times 10$	&	S	&	2.3	&	$-0.4 $ 	&	\nodata	&	\nodata	&	\nodata	&	\nodata	&	\nodata	&	\nodata	&	\nodata	&	2.88		&	18	\\
\enddata
\tablecomments{a - SNR types S: Shell, F: Filled centre and C: Composite. \\
b - Ages with a `*' were estimated using assumed parameters as described in Section \ref{Ages}.\\
c - The average radius, r$_{Avg}$ = $ \sqrt{r_{maj}S_{min}}$, where r$_{maj}$ is the semi-major axis and  r$_{min}$ is the semi-minor axis. \\
d - Units are $\times10^{-21}$ W m$^{-2}$ Hz$^{-1}$ Sr$^{-1}$. 
 }
\end{deluxetable*}
\end{turnpage}

\clearpage

\bibliography{References}{}
\bibliographystyle{aasjournal}



\end{document}